\documentclass[useAMS,usenatbib]{mn2e}
\usepackage{txfonts}
\usepackage{natbib}
\usepackage[all]{xy}
\usepackage[dvips]{graphicx}

\def\del#1{{}}

\newcommand{\ltsima}{$\; \buildrel < \over \sim \;$}
\newcommand{\lsim}{\lower.5ex\hbox{\ltsima}}
\newcommand{\gtsima}{$\; \buildrel > \over \sim \;$}
\newcommand{\gsim}{\lower.5ex\hbox{\gtsima}}
\newcommand{\bra}{\langle}
\newcommand{\ket}{\rangle}

\newcommand{\dd}{\mathrm{d}}

\title[Mixed iSW three-point functions]
{Mixed three-point correlation functions of the nonlinear integrated Sachs-Wolfe effect and their detectability}
\author[Bj{\"o}rn Malte Sch{\"a}fer]
{Bj{\"o}rn Malte Sch\"afer\thanks{e-mail: Bjoern.Malte.Schaefer@ias.u-psud.fr}\\
Institute of Cosmology and Gravitation, University of Portsmouth, Mercantile House, Hampshire Terrace, Portsmouth PO12EG, United Kingdom\\
Institut d'Astrophysique Spatiale, Universit{\'e} de Paris XI, b{\^a}timent 120-121, Centre universitaire d'Orsay, 91400 Orsay CEDEX, France}

\begin{document}
\pagerange{\pageref{firstpage}--\pageref{lastpage}}
\pubyear{2008}
\maketitle
\label{firstpage}

\begin{abstract}
In this paper I investigate the family of mixed three-point correlation functions $\bra\tau^q\gamma^{3-q}\ket$, $q=0,1,2$, between the integrated Sachs-Wolfe (iSW) temperature perturbation $\tau$ and the galaxy overdensity $\gamma$ as a tool for detecting the gravitational interaction of cosmic microwave background (CMB) photons with the potentials of nonlinearly evolving cosmological structures. Both the iSW-effect as well as the galaxy overdensity are derived in hyper-extended perturbation theory to second order and I emphasise the different parameter sensitivities of the linear and non-linear iSW-effect. I examine the configuration dependence of the relevant bispectra, quantify their sensitivities and discuss their degeneracies with respect to the cosmological parameters $\Omega_m$, $\sigma_8$, $h$ and the dark energy equation of state parameter $w$. I give detection sigificances for combining PLANCK CMB data and the DUNE galaxy sample, by using a quadratic approximation for the likelihood with $\Lambda$CDM as the fiducial cosmology: The combination of PLANCK with DUNE should be able to reach a cumulative signal to noise ratio of $\simeq0.6\sigma$ for the bispectrum $\bra\tau\gamma^2\ket$ up to $\ell=2000$, where the most important noise source are the primary CMB fluctuations.
\end{abstract}


\section{Introduction}
The integrated Sachs-Wolfe (iSW) effect \citep{1967ApJ...147...73S, 1994PhRvD..50..627H} is a secondary anisotropy in the cosmic microwave background (CMB) sky due to gravitational interaction of CMB photons with time-varying potentials in the cosmic large-scale structure (LSS) . The linear iSW-effect is an interesting observational channel because it directly measures the dark energy (DE) content of the universe due to its influence on the growth function. The linear iSW-effect has been detected with high statistical significance with different tracer objects  \citep{2003ApJ...598...97N, 2003ApJ...597L..89F, 2006PhRvD..74f3520G, 2007MNRAS.377.1085R}, and constraints on cosmological parameters such as $\Omega_m$ and the dark energy equation of state parameter $w$ can be derived \citep{1996PhRvL..76..575C, 2008arXiv0801.4380G, 2008arXiv0802.0983D}. 

On smaller angular scales one expects a contribution to the iSW-effect caused by gravitational interaction of the CMB photons with nonlinearly evolving cosmic structures, which is refered to as the Rees-Sciama (RS) effect \citep{rees_sciama_orig}. The angular power spectrum of the RS-effect has been derived in perturbation theory \citep{1990ApJ...355L...5M, 1996ApJ...463...15T, 1996ApJ...460..549S, 2002PhRvD..65h3518C, 2006MNRAS.369..425S}, and modelled on $n$-body simulations \citep{2006MNRAS.370.1849P, 2007A&A...476...83M}.

In this paper, I intend to derive mixed bispectra between the iSW-temperature perturbation and the density of tracer objects, as a tool of measuring the RS-effect caused by nonlinearly evolving cosmic structures, which give rise to non-Gaussian features in the CMB. I use hyper-extended perturbation theory for describing nonlinear growth and summarise the key equations related to cosmic structure formation in Sect.~\ref{sect_key}. In addition, I discuss the different parameter sensitivity of the RS-effect compared to the iSW-effect in this section. The mixed bispectra $\bra\gamma^3\ket$, $\bra\gamma^2\tau\ket$ and $\bra\gamma\tau^2\ket$ of the galaxy overdensity $\gamma$ and the iSW-temperature perturbation $\tau$ are derived in Sect.~\ref{sect_bispectra}, and I provide a discussion of their configuration dependence. I focus on these three bispectra because $\bra\tau^3\ket$ has already been shown to be very small \citep{1995ApJ...453....1M, 1999PhRvD..59j3001S, 2002PhRvD..65d3007V}, and because it would not be possible to distinguish primordial non-Gaussian features in the CMB from those induced by a secondary anisotropy without adding information about the LSS in form of the tracer density. In Sect.~\ref{sect_fisher}, I derive signal to noise ratios for the measurement of these bispectra for combining PLANCK\footnote{\tt http://www.rssd.esa.int/index.php?project=Planck} CMB data with the main galaxy sample of the Dark UNiverse Explorer\footnote{\tt http://www.dune-mission.net/} (DUNE), which covers half of the sky and contains about $3\times10^9$ objects. Then I continue with a discussion of the parameter sensitivity of the mixed bispectra (the relevant parameters being $\Omega_m$, $\sigma_8$, $w$ and $h$) and summarise my main results in Sect.~\ref{sect_summary}

At this point I should emphasise that I consider only idealised measurements, i.e. the PLANCK CMB-observation is tainted with a Gaussian noise component and a Gaussian beam, and the galaxy surveys are assumed to have a Poissonian noise component in the galaxy number density. A quantification of systematic effects, such as unresolved point sources or the kinetic Sunyaev-Zel'dovich effect, which both are equally associated with overdensities in the LSS, is beyond the scope of this paper. 

The cosmological model used is the spatially flat $\Lambda$CDM cosmology with adiabatic initial conditions. Parameter values are $\Omega_m=0.25$, $\sigma_8=0.8$ and $n_s=1$, with a constant equation of state $w\equiv -1$ for the dark energy fluid. The Hubble-constant has the value $H_0=100\:h\:\mathrm{km}/\mathrm{s}/\mathrm{Mpc}$. For simplicity, the DUNE galaxy sample is assumed to have a non-evolving bias of unity, $b=1$.

\section{key formulae}\label{sect_key}

\subsection{Dark energy cosmologies}
In spatially flat dark energy cosmologies with the matter density parameter $\Omega_m$, the Hubble function $H(a)=\dd\ln a/\dd t$ is given by
\begin{equation}
\frac{H^2(a)}{H_0^2} = \frac{\Omega_m}{a^{3}} + (1-\Omega_m)\exp\left(3\int_a^1\dd\ln a\:(1+w(a))\right),
\end{equation}
with the dark energy equation of state $w(a)$. The value $w\equiv -1$ corresponds to the cosmological constant $\Lambda$. The relation between comoving distance $\chi$ and scale factor $a$ is given by
\begin{equation}
\chi = c\int_a^1\dd a\:\frac{1}{a^2 H(a)},
\end{equation}
with the speed of light $c$.

\subsection{Linear structure formation and power spectra}
For the linear CDM density power spectrum $P(k)$, defined from the fluctuation amplitude of the density field, $\bra\delta(\bmath{k})\delta(\bmath{k}^\prime)\ket=(2\pi)^3\delta_D(\bmath{k}+\bmath{k}^\prime)P(k)$, I use the the ansatz
\begin{equation}
P(k)\propto k^{n_s}T^2(k),
\end{equation}
with the transfer function $T(k)$. The transfer function is approximated with the polynomial fit proposed by \citep{1986ApJ...304...15B},
\begin{displaymath}
T(q) = \frac{\ln(1+2.34q)}{2.34q}\left(1+3.89q+(16.1q)^2+(5.46q)^3+(6.71q)^4\right)^{-\frac{1}{4}}.
\label{eqn_cdm_transfer}
\end{displaymath}
The wave vector $k=q\Gamma$ is rescaled with the shape parameter $\Gamma$ \citep{1995ApJS..100..281S},
\begin{equation}
\Gamma=\Omega_m h\exp\left(-\Omega_b\left(1+\frac{\sqrt{2h}}{\Omega_m}\right)\right).
\end{equation}
The spectrum $P(k)$ is normalised to the variance $\sigma_8$ on the scale $R=8~\mathrm{Mpc}/h$,
\begin{equation}
\sigma^2_R = \frac{1}{2\pi^2}\int\dd k\:k^2 P(k) W^2(kR),
\end{equation}
with a Fourier transformed spherical top hat filter function, $W(x)=3j_1(x)/x$. $j_\ell(x)$ is the spherical Bessel function of the first kind of order $\ell$  \citep{1972hmf..book.....A}. The homogeneous growth of the density field in the linear regime, $\delta(\bmath{x},a)=D_+(a)\delta(\bmath{x},a=1)$, is described with the growth function $D_+(a)$, which results from solving the growth equation \citep{1998ApJ...508..483W, 1997PhRvD..56.4439T, 2003MNRAS.346..573L},
\begin{equation}
\frac{\dd^2}{\dd a^2}D_+(a) + \frac{1}{a}\left(3+\frac{\dd\ln H}{\dd\ln a}\right)\frac{\dd}{\dd a}D_+(a) = 
\frac{3}{2a^2}\Omega_m(a) D_+(a),
\label{eqn_growth}
\end{equation}
and which assumes the simple solution $D_+(a)=a$ in the SCDM cosmology, where $\Omega_m\equiv 1$ and $H(a)\propto a^{-3/2}$.

\subsection{Galaxy biasing}
The fractional perturbation $\Delta n/\bra n\ket$ of the mean number density $\bra n\ket$ of galaxies is related to the overdensity $\delta$ of dark matter. I use the phenomenological relation
\begin{equation}
\frac{\Delta n}{\bra n\ket}=b\frac{\Delta\rho}{\bra\rho\ket} = b\delta,
\end{equation}
with a constant bias parameter $b$. There exist more elaborate nonlinear biasing models, even including time evolution (which itself can be parameterised with e.g. $b(a)=b_0 + (1-a)b_a$), but for simplicity, I will work with a constant linear biasing model.

\subsection{Linear and nonlinear iSW-effects}\label{sect_isw_time_evolution}
The iSW-effect is caused by gravitational interaction of a CMB photon with a time-evolving potential $\Phi$. The fractional perturbation $\tau$ of the CMB temperature $T_\mathrm{CMB}$ is given by \citep{1967ApJ...147...73S}
\begin{equation}
\tau 
= \frac{\Delta T}{T_\mathrm{CMB}} 
\equiv -\frac{2}{c^2}\int\dd\eta\: \frac{\partial\Phi}{\partial\eta} 
= \frac{2}{c^3}\int_0^{\chi_H}\dd\chi\: a^2 H(a) \frac{\partial\Phi}{\partial a},
\label{eqn_sachs_wolfe}
\end{equation}
where $\eta$ denotes the conformal time, and the integration is extended to the horizon distance $\chi_H$. In the last step, I have replaced the integration variable by the comoving distance $\chi$, which is related to the conformal time by $\dd\chi = -c\dd\eta = -c\dd t/a$, and the time derivative of the growth function has been rewritten in terms of the scale factor $a$, using the definition of the Hubble function $\dd a/\dd t = aH(a)$, with the cosmic time $t$. The gravitational potential $\Phi$ follows from the Poisson equation in the comoving frame, where the Newton's constant $G$ is replaced with the critical density $\rho_\mathrm{crit}=3H_0^2/(8\pi G)$,
\begin{equation}
\Delta\Phi = \frac{3H_0^2\Omega_m}{2a}\delta.
\end{equation}
Substitution yields a line of sight expression for the linear iSW-effect $\tau^{(1)}$ (integrating along a straight line and using the flat-sky approximation), sourced by the linear density field $\delta^{(1)}$,
\begin{equation}
\tau^{(1)} = 
\frac{3\Omega_m}{c}\int_0^{\chi_H}\dd\chi\: 
a^2 H(a)\:\frac{\dd}{\dd a}\left(\frac{D_+}{a}\right)\:\frac{\Delta^{-1}}{d_H^2}\delta^{(1)},
\end{equation}
with the inverse (dimensionless) Laplace operator $\Delta^{-1}/d_H^2$ solving for the (dimensionless) potential $\varphi$,
\begin{equation}
\varphi^{(i)} \equiv \frac{\Delta^{-1}}{d_H^2}\delta^{(i)},
\end{equation}
with the Hubble distance $d_H=c/H_0$. Extending this expression to include a nonlinear correction $\delta^{(2)}$ to the density field, $\delta=D_+(a)\delta^{(1)}+D_+^2(a)\delta^{(2)}$, gives a contribution $\tau^{(2)}$ from nonlinear structure formation,
\begin{equation}
\tau^{(2)} = 
\frac{3\Omega_m}{c}\int_0^{\chi_H}\dd\chi\:
a^2 H(a)\:\frac{\dd}{\dd a}\left(\frac{D_+^2}{a}\right)\:\frac{\Delta^{-1}}{d_H^2}\delta^{(2)}.
\label{eqn_rees_sciama}
\end{equation}
This second order effect \citep{rees_sciama_orig} has in fact different parameter dependences, as illustrated by a simple example: The iSW-effect vanishes in the SCDM-model ($\Omega_m=1$), because $D_+(a)=a$, and is nonzero in models with dark energy, because the growth function $D_+(a)$ scales slower at low redshifts. On the contrary, the RS-effect is present in the SCDM-cosmology, as it measures $\dd(D_+^2/a)/\dd a$ instead of $\dd(D_+/a)/\dd a$, and is at low redshifts always smaller in dark energy cosmologies compared to SCDM. 

\begin{figure}
\resizebox{\hsize}{!}{\includegraphics{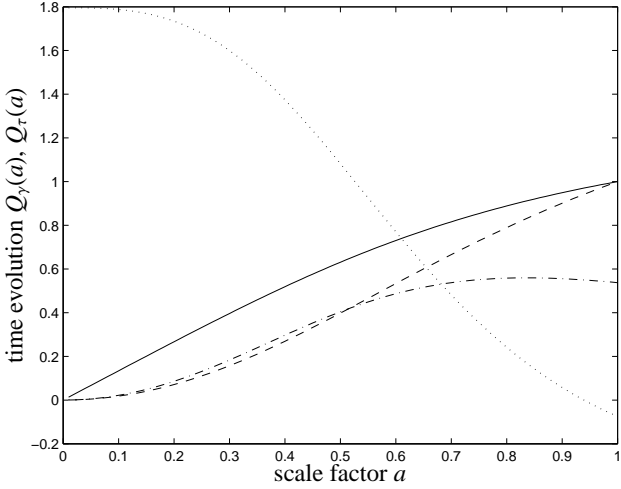}}
\caption{Time evolution $Q_\gamma(a)$ and $Q_\tau(a)$ of the source fields as a function of scale factor $a$: density field in first order $\delta^{(1)}$ (solid line), density field in second order $\delta^{(2)}$ (dashed line), potential derivative in first order $\dot{\varphi}^{(1)}$ (dash-dotted line) and potential derivative in second order $\dot{\varphi}^{(2)}$ (dotted line).}
\label{fig_source_evolution}
\end{figure}

The line of sight expression for the projected galaxy overdensity $\gamma$ in first and second order is given by
\begin{eqnarray}
\gamma^{(1)} & = & \int_0^{\chi_H}\dd\chi\: p(z)\frac{\dd z}{\dd\chi}\: D_+ b\:\delta^{(1)}, \\
\label{eqn_density_1}
\gamma^{(2)} & = & \int_0^{\chi_H}\dd\chi\: p(z)\frac{\dd z}{\dd\chi}\: D_+^2 b\:\delta^{(2)},
\label{eqn_density_2}
\end{eqnarray}
with the redshift distribution $p(z)\dd z$ and the linear bias parameter $b$. $p(z)\dd z$ is approximated by \citep{1995MNRAS.277....1S},
\begin{equation}
p(z)\dd z = p_0\left(\frac{z}{z_0}\right)^2\exp\left(-\left(\frac{z}{z_0}\right)^\beta\right)\dd z
\quad\mathrm{with}\quad \frac{1}{p_0}=\frac{z_0}{\beta}\Gamma\left(\frac{3}{\beta}\right).
\end{equation}
Fig.~\ref{fig_source_evolution} shows the time-evolution $Q_\gamma(a)$ of the the source fields $\delta^{(i)}$,
\begin{equation}
Q_\gamma(a) = 
\left\{
\begin{array}{ll}
D_+(a), 								& \mathrm{for}~\delta^{(1)}\\
D_+^2(a), 								& \mathrm{for}~\delta^{(2)}\\
\end{array}
\right.
\label{eqn_time_evolution_delta}
\end{equation}
and $Q_\tau(a)$ of $\varphi^{(i)}$ which result in the galaxy overdensity $\gamma$ and for the iSW-effect $\tau$ by projection, respectively:
\begin{equation}
Q_\tau(a) = 
\left\{
\begin{array}{ll}
\frac{\dd}{\dd a}\frac{D_+}{a}, 		& \mathrm{for}~\varphi^{(1)}\\
\frac{\dd}{\dd a}\frac{D_+^2}{a}, 	& \mathrm{for}~\varphi^{(2)}.
\end{array}
\right.
\label{eqn_time_evolution_phi}
\end{equation}

\begin{figure}
\resizebox{\hsize}{!}{\includegraphics{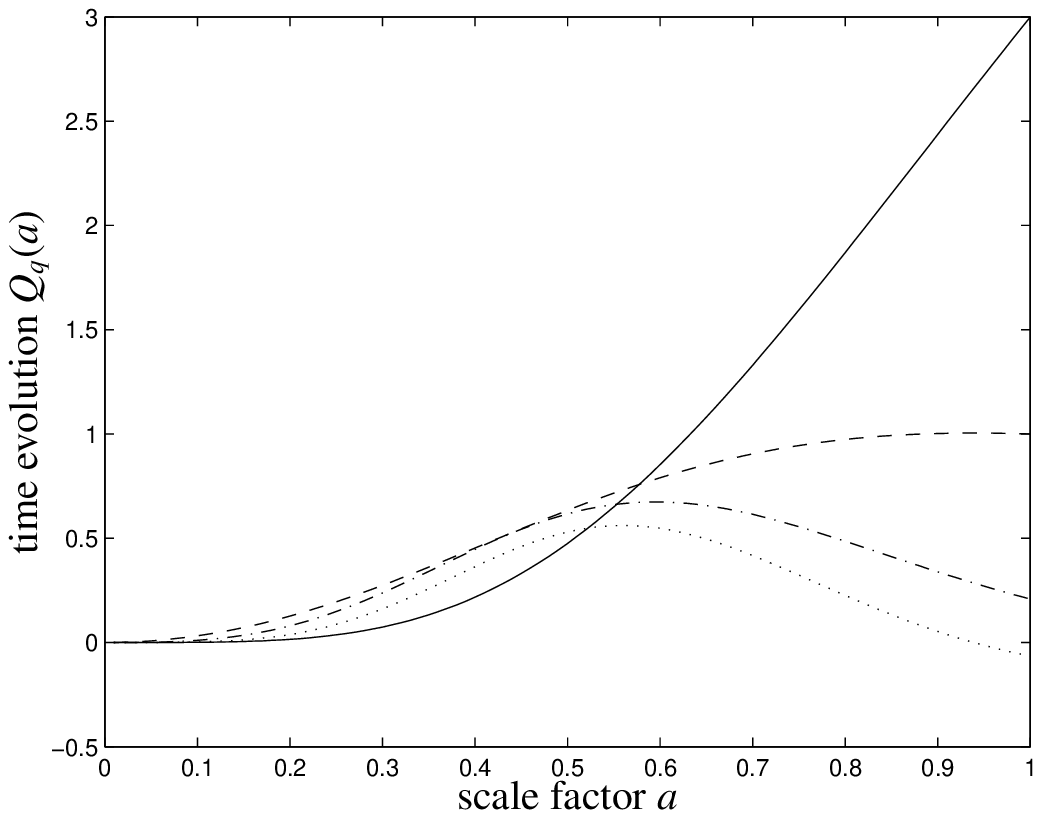}}
\caption{Time evolution $Q_q(a)$ of the bispectra $\bra \tau^q\gamma^{3-q}\ket$, for $\bra\gamma^3\ket$ (solid line), $\bra\tau\gamma^2\ket$ (dashed line) $\bra\tau^2\gamma\ket$ (dash-dotted line) and $\bra\tau^3\ket$ (dotted line), in second order perturbation theory, as a function of scale factor $a$.}
\label{fig_time_evolution}
\end{figure}

I define $Q_q(a)$ to be the time evolution of the bispectrum $\bra\varphi^q\delta^{3-q}\ket$ of the source fields, which acquires terms from first and second order growth (anticipating results from Sect.~\ref{sect_perturbation}),
\begin{equation}
Q_q(a) =
\left\{
\begin{array}{ll}
3 D_+^4, 	
	&\mathrm{for}~\bra\delta^3\ket \\
D_+^2\left(\frac{\dd}{\dd a}\frac{D_+^2}{a}\right) + 2D_+^3(a)\left(\frac{\dd}{\dd a}\frac{D_+}{a}\right),
	&\mathrm{for}~\bra\delta^2\varphi\ket\\
2D_+\left(\frac{\dd}{\dd a}\frac{D_+^2}{a}\right)\left(\frac{\dd}{\dd a}\frac{D_+}{a}\right) + D_+^2\left(\frac{\dd}{\dd a}\frac{D_+}{a}\right)^2,
	&\mathrm{for}~\bra\delta\varphi^2\ket\\
3 \left(\frac{\dd}{\dd a}\frac{D_+}{a}\right)^2\left(\frac{\dd}{\dd a}\frac{D_+^2}{a}\right)
	&\mathrm{for}~\bra\varphi^3\ket.
\end{array}
\right.
\end{equation}
It is apparent how the different terms in the first and second order time evolution of the density and the potential field affect the time evolution of the bispectrum $\bra\varphi^q\delta^{3-q}\ket$. Up to this moment I have included the iSW-bispectrum $B_3=\bra\tau^3\ket$ for completeness. I would like to point out that it is not possible to distinguish non-Gaussian features imprinted into the CMB by the iSW-effect from primordial non-Gaussianities, due to the achromaticity of the iSW-effect. For this reason, I will only consider the bispectra $\bra\gamma^3\ket$, $\bra\tau\gamma^2\ket$ and $\bra\tau^2\gamma\ket$, which can be measured in cross-correlation with a tracer field $\gamma$. Fig.~\ref{fig_time_evolution} shows the time-evolution $Q_q(a)$ of the 3-point correlation functions $\bra\varphi^q\delta^{3-q}\ket$ as a function of scale factor $a$.

\subsection{Weighting functions}
\begin{figure}
\resizebox{\hsize}{!}{\includegraphics{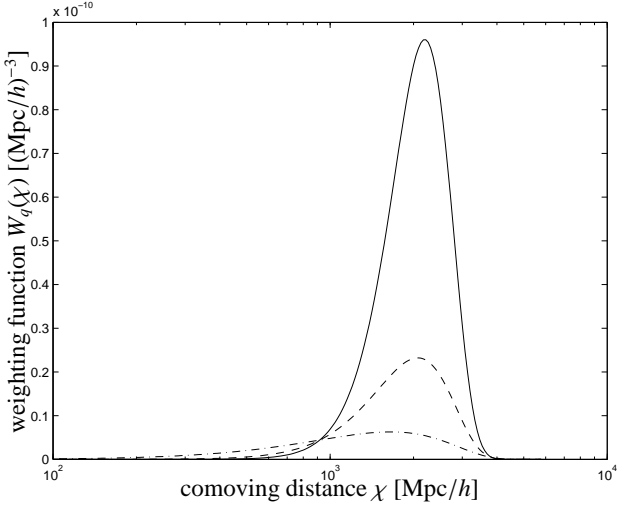}}
\caption{Line-of-sight weighting functions $W_q(\chi)$ for the projection of the bispectra $\bra\tau^q\gamma^{3-q}\ket$, for $\bra \gamma^3\ket$ (solid line), $\bra\tau \gamma^2\ket$ (dashed line) and $\bra\tau^2 \gamma\ket$ (dash-dotted line), as a function of comoving distance $\chi$, for the redshift distribution of the DUNE galaxy sample.}
\label{fig_weighting}
\end{figure}

For composing the weighting functions $W_q(\chi)$ needed in the projection~(\ref{eqn_limber}), one can read off the weightings 
\begin{eqnarray}
W_\tau(\chi) & = & \frac{3\Omega_m}{c}a^2 H(a),\label{eqn_w_tau}\\
W_\gamma(\chi) & = & b p(z)\frac{\dd z}{\dd\chi} = \frac{b}{c} p(z) H(a) \label{eqn_w_gamma}
\end{eqnarray}
from eqns.~(\ref{eqn_sachs_wolfe}) and~(\ref{eqn_density_1}), which both have units of inverse $(\mathrm{Mpc}/h)$. With $W_\tau(\chi)$ and $W_\gamma(\chi)$, the weighting functions $W_q(\chi)$ used for carrying out the projection of the soure field bispectra $\bra\varphi^q\delta^{3-q}\ket$ to the angular bispectra $\bra\tau^q\gamma^{3-q}\ket$ can be written down,
\begin{equation}
W_q(\chi)
= W_\tau^q(\chi)W_\gamma^{3-q}(\chi)
=\left\{
\begin{array}{ll}
W_\gamma^3(\chi), 				&\mathrm{for}~\bra\gamma^3\ket \\
W_\gamma^2(\chi)W_\tau(\chi),	&\mathrm{for}~\bra\gamma^2\tau\ket\\
W_\gamma(\chi)W_\tau^2(\chi),	&\mathrm{for}~\bra\gamma\tau^2\ket\\
W_\tau^3(\chi),					&\mathrm{for}~\bra\tau^3\ket.
\end{array}
\right.
\end{equation}
The weighting functions $W_q(\chi)$ are depicted in Fig.~\ref{fig_weighting}: Common to all $W_q(\chi)$ is a broad peak between $1~\mathrm{Gpc}/h$ and $4~\mathrm{Gpc}/h$, corresponding to the maximum of the galaxy redshift distribution.

\section{Mixed 3-point functions}\label{sect_bispectra}

\subsection{Bispectra}
Bispectra are a tool for quantifying non-Gaussianities which arise in the course of structure formation when nonlinearities in the structure formation equations set in. Bispectra are the Fourier analogue of three-point correlation functions, i.e. for the density field $\delta$ the relation 
\begin{equation}
\bra\delta(\bmath{k}_1)\delta(\bmath{k}_2)\delta(\bmath{k}_3)\ket = 
(2\pi)^3\delta_D(\bmath{k}_1+\bmath{k}_2+\bmath{k}_3) 
B_{\delta}(\bmath{k}_1,\bmath{k}_2,\bmath{k}_3)
\end{equation}
defines the bispectrum $B_\delta$ via the three-point variance of the Fourier-transformed density field. Equivalent formulae apply to angular bispectrum $B_\gamma$ of the projected density field $\gamma$ 
\begin{equation}
\bra\gamma(\bmath{\ell}_1)\gamma(\bmath{\ell}_2)\gamma(\bmath{\ell}_3)\ket = 
(2\pi)^3 \delta_D(\bmath{\ell}_1+\bmath{\ell}_2+\bmath{\ell}_3) 
B_\gamma(\bmath{\ell}_1,\bmath{\ell}_2,\bmath{\ell}_3),
\end{equation}
in the flat-sky approximation, with angular wave vectors $\bmath{\ell}$.

\subsection{Notation}
I introduce a notation reminiscent of weak lensing tomography: The bispectra $B_{(abc)}(\ell_1,\ell_2,\ell_3)$ are indexed by the variables $a,b,c$, which assume the value one for the iSW-field $\tau$ and zero for the galaxy overdensity $\gamma$. This is a way to distinguish between e.g. the bispectra $B_{(100)}(\ell_1,\ell_2,\ell_3)=\bra\tau(\ell_1)\gamma(\ell_2)\gamma(\ell_3)\ket$, $B_{(010)}(\ell_1,\ell_2,\ell_3)=\bra\gamma(\ell_1)\tau(\ell_2)\gamma(\ell_3)\ket$ and $B_{(001)}(\ell_1,\ell_2,\ell_3)=\bra\gamma(\ell_1)\gamma(\ell_2)\tau(\ell_3)\ket$, assuming $\sum_p\bmath{\ell}_p=0$. These bisepctra are not equal (meaning the triangles can not be mapped onto each other for general configurations), unlike permutations in weak lensing tomography bin numbers. Most of the signal in an actual measurement, however, will come from general configurations and not from isosceles or equilateral configurations, for which these bispectra would be in fact identical. Nevertheless, all permutations with fixed $q=a+b+c$ share the same time evolution $Q_q(a)$ and the same weighting function $W_q(\chi)$. 

I relate the source field bispectra $B_{(abc)}$ to the bispectrum of the density field $B_{(000)}$ with the formula
\begin{equation}
B_{(abc)}(\bmath{k}_1,\bmath{k}_2,\bmath{k}_3) = 
\left(\frac{-1}{d_H^2}\right)^{a+b+c}
\frac{1}{k_1^{2a}k_2^{2b}k_3^{2c}}\:
B_{(000)}(\bmath{k}_1,\bmath{k}_2,\bmath{k}_3),
\end{equation}
and average over all combinations with fixed $q$ for visualisation purposes: The bispectra $B_q(\ell_1,\ell_2,\ell_3)$ depicted in Figs.~\ref{fig_bispectrum_equilateral} through~\ref{fig_bispectrum_configuration2} will be averages over all permutations, i.e.
\begin{equation}
B_q(\ell_1,\ell_2,\ell_3) = \frac{(-1)^q}{{3 \choose q}}\:\sum_{a,b,c \atop a+b+c=q}B_{(abc)}(\ell_1,\ell_2,\ell_3),
\end{equation}
with the binomial coefficient ${3\choose q}=\frac{3!}{q!(3-n)!}$ counting the permutations with a fixed number of $q$ iSW-fields $\tau$ and $3-q$ galaxy overdensity fields $\gamma$. In total, there are three permutations for $B_1$ and $B_2$, and one permutation for $B_0$ and $B_3$. The occurence of first and second order terms of the respective fields $\delta$, $\varphi$ in the perturbative expansion determines the time evolution of the bispectrum $\bra\varphi^q\delta^{3-q}\ket$ and lead to the equations in Sect.~\ref{sect_isw_time_evolution}.

\subsection{Nonlinear structure formation and perturbation theory}\label{sect_perturbation}
In order to carry out the evaluation of the bispectra in perturbation theory I use the reformulation of the iSW-effect outlined above such that the souce fields $\delta^{(1)}$ and $\varphi^{(1)}$ are dimensionless, as well as their respective first and second order time evolution $Q_\tau(a)$, $Q_\gamma(a)$ and their second order perturbations $\delta^{(2)}$ and $\varphi^{(2)}$. The weighting functions $W_\tau(\chi)$ and $W_\gamma(\chi)$ used for carrying out the line of sight projection will have units of inverse $\mathrm{Mpc}/h$, such that the units are consistent in the Limber-projection of bispectra, and that the resulting angular bispectra $\bra\tau^q\gamma^{3-q}\ket$ are dimensionless. Using the formalism of \citet{1996ApJ...460..549S}, I separate the time evolution and the mode-coupling of the source fields.

The first order contribution to the bispectrum $B_0(\bmath{k}_1,\bmath{k}_2,\bmath{k}_3)=\bra\delta(\bmath{k}_1)\delta(\bmath{k}_2)\delta(\bmath{k}_3)\ket$ of the density field from nonlinear structure formation is can be expanded into a product of power spectra $P_{\delta\delta}(k)$ \citep[I will follow][and use the nonlinear power spectrum for $P_{\delta\delta}(k)$]{2003MNRAS.340..580T} and is given by \citep{1984ApJ...277L...5F, 1984ApJ...279..499F}:
\begin{equation}
B_0(\bmath{k}_1,\bmath{k}_2,\bmath{k}_3) = 
\sum_{{(i,j)\in\left\{1,2,3\right\} \atop i\neq j}}
M(\bmath{k}_i,\bmath{k}_j)P_{\delta\delta}(k_i)P_{\delta\delta}(k_j),
\end{equation}
with the mode coupling functions of classical perturbation theory,
\begin{equation}
M(\bmath{k}_i,\bmath{k}_j) =
\frac{10}{7} + \left(\frac{k_i}{k_j}+\frac{k_j}{k_i}\right)x + \frac{4}{7}x^2,
\end{equation}
where $x =\bmath{k}_i\bmath{k}_j/\left(k_ik_j\right)$ denotes the cosine of the angle between $\bmath{k}_i$ and $\bmath{k}_j$.  In hyper-extended perturbation theory \citep{1999ApJ...520...35S, 2001MNRAS.325.1312S}, the mode coupling function is replaced by:
\begin{equation}
M(\bmath{k}_i,\bmath{k}_j) =
\frac{10}{7} a(k_i)a(k_j) 
+ b(k_i)b(k_j)\left(\frac{k_i}{k_j}+\frac{k_j}{k_i}\right)x
+ \frac{4}{7}c(k_i)c(k_j)x^2.
\end{equation}
The coefficients $a(k)$, $b(k)$ and $c(k)$ are given by :
\begin{eqnarray}
a(k) & = & \frac{1+\sigma_8^{-0.2}(z)\sqrt{0.7Q(n)}\:(q/4)^{n+3.5}}{1+(q/4)^{n+3.5}},\label{eqn_hept_a}\\
b(k) & = & \frac{1+0.4(n+3)q^{n+3}}{1+q^{n+3.5}},\label{eqn_hept_b}\\
c(k) & = & \frac{1+\frac{4.5}{1.5+(n+3)^4}\:(2q)^{n+3}}{1+(2q)^{n+3.5}},\label{eqn_hept_c}
\end{eqnarray}
where the time evolution of the fluctuation amplitude is given by the linear growth function, $\sigma_8(z)=D_+(z)\sigma_8$. In eqns.~(\ref{eqn_hept_a}), (\ref{eqn_hept_b}) and~(\ref{eqn_hept_c}), the wave vectors $k$ are expressed in units of the nonlinear wave number $k_\mathrm{NL}$, $q\equiv k/k_{\mathrm{NL}}$. The nonlinear wave number at scale factor $a$ is given by the scale at which the variance $\sigma^2$ of the density fluctuations becomes unity,
\begin{equation}
\sigma^2=\int_0^{k_\mathrm{NL}}\dd^3k\:D_+^2(z)P(k)=1\rightarrow
4\pi k_\mathrm{NL}^3D_+^2(z)P(k_{\mathrm{NL}})=1.
\end{equation}
The logarithmic slope of the linear power spectrum,
\begin{equation}
n(k) = \frac{\dd\ln P(k)}{\dd\ln k}
\end{equation}
can be directly derived with form of the transfer function $T(k)$ given in eqn.~(\ref{eqn_cdm_transfer}), and is used for determining the saturation parameter $Q(n)$, which is defined as the logarithmic slope $n$ of the linear CDM spectrum,
\begin{equation}
Q(n) = \frac{4-2^n}{1+2^{n+1}}.
\end{equation}
I use the functional form derived by \citet{2003MNRAS.341.1311S} for the nonlinear CDM spectrum $P_{\delta\delta}(k)$ and its slow time evolution, parameterised with $\Omega_m(a)$.

\subsection{Limber projection}
In order to relate the source field bispectra $\bra\varphi^q\delta^{3-q}\ket$ to the angular bispectra $\bra\tau^q\gamma^{3-q}\ket$ of the observables $\tau$ and $\gamma$, I carry out a Limber-style projection \citep{1954ApJ...119..655L},
\begin{equation}
B_q(\bmath{\ell}_1,\bmath{\ell}_2,\bmath{\ell}_3) = 
\int_0^{\chi_H}\dd\chi\: \frac{1}{\chi^4}\: W_q(\chi) Q_q(\chi)\: B_q(\bmath{k}_1,\bmath{k}_2,\bmath{k}_3),
\label{eqn_limber}
\end{equation}
for which I adopt the flat-sky approximation, which is justified as the nonlinear iSW-effect is a small-scale phenomenon. The relation between the wave vectors $\bmath{k}_p$ and the multipole vectors $\bmath{\ell}_p$ is given by $\bmath{k}_p=\bmath{\ell}_p/\chi$, $p=1,2,3$. The spherical bispectrum $B_q(\ell_1,\ell_2,\ell_3)$ is related to the flat-sky bispectrum $B_q(\bmath{\ell}_1,\bmath{\ell}_2,\bmath{\ell}_3)$ via \citep{1991ApJ...380....1M,1992ApJ...388..272K} 
\begin{equation}
B_q(\ell_1,\ell_2,\ell_3) = 
\left(
\begin{array}{ccc}
\ell_1 	& \ell_2 	& \ell_3 \\
0	 	& 0 			& 0
\end{array}
\right)
\sqrt{\frac{\prod_{p=1}^3(2\ell_p+1)}{4\pi}}
B_q(\bmath{\ell}_1,\bmath{\ell}_2,\bmath{\ell}_3),
\end{equation}
where
\begin{equation}
\left(\begin{array}{ccc}\ell_1 & \ell_2 & \ell_3\\ 0 & 0 & 0\end{array}\right)^2 = 
\frac{1}{2}\int_{-1}^{+1}\dd x\: P_{\ell_1}(x)P_{\ell_2}(x)P_{\ell_3}(x),
\end{equation}
$x=\cos\theta$, denotes the Wigner-$3j$ symbol, which results from integrating over three Legendre polynomials $P_\ell(x)$ \citep{1972hmf..book.....A}. The Wigner-$3j$ cancels configurations which do not satisfy the triangle inequality $\left|\ell_i-\ell_j\right|\leq\ell_k\leq\ell_i+\ell_l$. The factorials arising in the evaluation of the Wigner-$3j$ symbol are computed using the Stirling-approximation for the $\Gamma$-function, $\Gamma(n) = (n-1)!$,
\begin{equation}
\Gamma(x)\simeq\sqrt{2\pi}\exp(-x)x^{x-\frac{1}{2}},
\end{equation}
\citep{1972hmf..book.....A}, which is valid for large $x$, $x\gg 1$ and gives sufficient accuracy (roughly 0.4\% in the relevant $\ell$-range) for the purpose of this study.

\subsection{Angular bispectra}
\begin{figure}
\resizebox{\hsize}{!}{\includegraphics{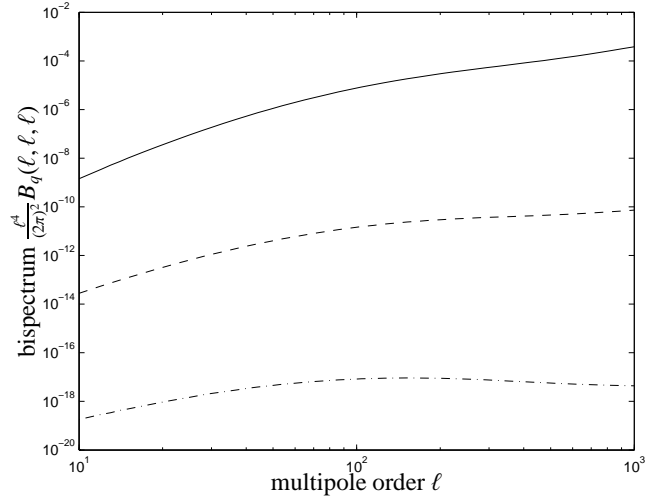}}
\caption{Angular equilateral bispectra $\bra \gamma^3\ket$ (solid line), $\bra\tau \gamma^2\ket$ (dashed line) and $\bra\tau^2 \gamma\ket$ (dash-dotted line), as a function of multipole order $\ell$, for the redshift distribution of the DUNE galaxy sample.}
\label{fig_bispectrum_equilateral}
\end{figure}

This section gives an overview over the different configuration and scale dependences of the the angular bispectra $\bra\tau^q\gamma^{3-q}\ket$: The equilateral angular bispectrum $B_q(\ell,\ell,\ell)$ corresponding to the 3-point correlation functions $\bra\tau^q\gamma^{3-q}\ket$ are shown in Fig.~\ref{fig_bispectrum_equilateral}. There is clearly a hierarchy in the bispectra, with $\bra\gamma^3\ket$ attaining the largest values, followed by $\bra\gamma^2\tau\ket$ and $\bra\gamma\tau^2\ket$, separated by two orders of magnitude on large scales and up to seven orders of magnitude on small scales. The shape of the bispectra is determined by the mode-coupling functions $M(\bmath{k}_i,\bmath{k}_j)$ and the $k^{-2}$-factors in perturbation theory.

\begin{figure}
\resizebox{\hsize}{!}{\includegraphics{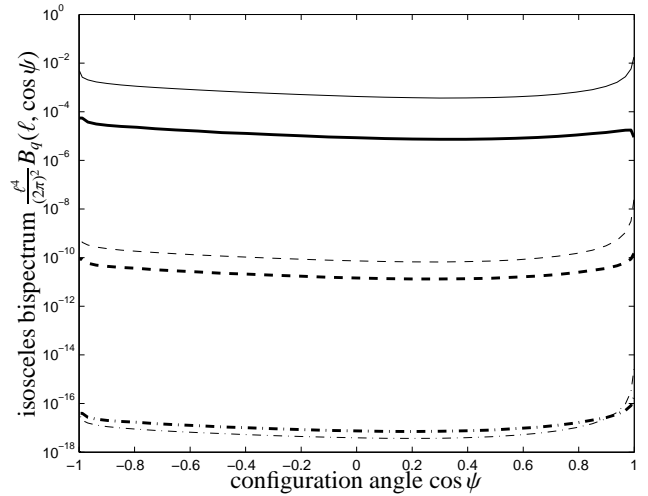}}
\caption{Angular isosceles bispectra $\bra \gamma^3\ket$ (solid line), $\bra\tau \gamma^2\ket$ (dashed line) and $\bra\tau^2 \gamma\ket$ (dash-dotted line), as a function of angle $\cos\psi$, for multipole orders $\ell=10^2$ (thick lines) and $\ell=10^3$ (thin lines), for the redshift distribution of the DUNE galaxy sample.}
\label{fig_bispectrum_isosceles}
\end{figure}

Apart from a clear scale dependence, the angular bispectrum $B_q(\ell,\cos\psi)$ corresponding to the 3-point correlation functions $\bra\tau^q\gamma^{3-q}\ket$ exhibit a different configuration dependence as well: Isosceles bispectra are shown in Fig.~\ref{fig_bispectrum_isosceles}, as a function of the opening angle $\psi$ of the triangle. Typical variations of the bispectra with configuration on a fixed angular scale amount to two orders of magnitude. The configuration dependence itself changes with angular scale, as the bispectra plotted for $\ell=100$ and for $\ell=1000$ show different dependences on $\psi$.

\begin{figure}
\resizebox{\hsize}{!}{\includegraphics{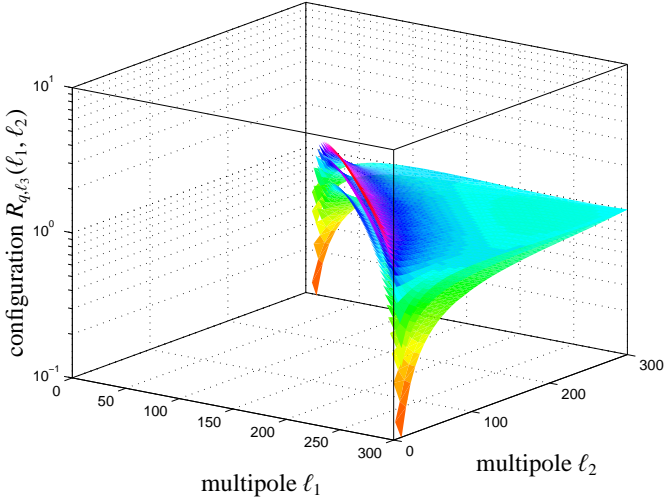}}
\caption{Configuration dependence $R_{q,\ell_3}(\ell_1,\ell_2)$ of the angular bispectra, $\bra \gamma^3\ket$ (bottom plane), $\bra\tau \gamma^2\ket$ (centre plane), $\bra\tau^2 \gamma\ket$ (top plane), as a function of $\ell_1$ and $\ell_2$ with $\ell_3=300$, for the redshift distribution of the DUNE galaxy sample. The empty region to the left would contain invalid configurations violating the triangle inequality $\left|\ell_1-\ell_2\right|\leq\ell_3\leq\ell_1+\ell_2$.}
\label{fig_bispectrum_configuration1}
\end{figure}

\begin{figure}
\resizebox{\hsize}{!}{\includegraphics{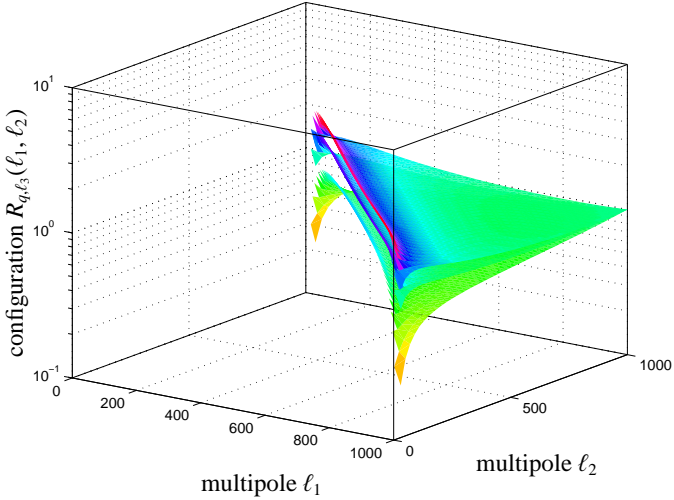}}
\caption{Configuration dependence $R_{q,\ell_3}(\ell_1,\ell_2)$ of the angular bispectra, $\bra \gamma^3\ket$ (bottom plane), $\bra\tau \gamma^2\ket$ (centre plane), $\bra\tau^2 \gamma\ket$ (top plane), as a function of $\ell_1$ and $\ell_2$ with $\ell_3=1000$, for the redshift distribution of the DUNE galaxy sample. The empty region to the left would contain invalid configurations violating the triangle inequality $\left|\ell_1-\ell_2\right|\leq\ell_3\leq\ell_1+\ell_2$.}
\label{fig_bispectrum_configuration2}
\end{figure}

The configuration dependence of the bispectra $\bra\tau^q\gamma^{3-q}\ket$ is shown for $\ell_3=300$ and $\ell_3=1000$ in Figs.~\ref{fig_bispectrum_configuration1} and~\ref{fig_bispectrum_configuration2}. Specifically, I plot the configuration space variable \citep{2002PhR...372....1C}
\begin{equation}
R_{q,\ell_3}(\ell_1,\ell_2) = 
\frac{\ell_1\ell_2}{\ell_3^2}
\sqrt{\left|\frac{B_q(\ell_1,\ell_2,\ell_3)}{B_q(\ell_3,\ell_3,\ell_3)}\right|},
\end{equation}
which shows the dependence of the bispectrum $B_q$ on $(\ell_1,\ell_2)$ with $\ell_3$ fixed, in comparison to the equilateral configuration. The reason for this behaviour are the different power spectra of the density and potential fields, the latter having a much larger correlation length due to the $k^{-2}$-factor. 

\subsection{Amplitude of the nonlinearities}\label{sect_amp_nonlinear}
The overall amplitude of the nonlinearities probed by the iSW-effect can be quantified by the ratio $\alpha$ between the squared bispectrum (e.g. for the equilateral configuration) and the corresponding power spectrum to the third power, 
\begin{eqnarray}
\alpha_\tau(\ell) 		& = & \left(\frac{B_{3}^2(\ell,\ell,\ell)}{C_{\tau\tau}^3(\ell)}\right)^\frac{1}{6},
\label{eqn_alpha_tau}\\
\alpha_\gamma(\ell)	& = & \left(\frac{B_{0}^2(\ell,\ell,\ell)}{C_{\gamma\gamma}^3(\ell)}\right)^\frac{1}{6}
\label{eqn_alpha_gamma}
\end{eqnarray}
for which I obtain values of $\alpha_\tau = 1.42\times10^{-3}, 1.18\times10^{-2}, 9.87\times10^{-2}$ and in comparison $\alpha_\gamma = 1.78\times10^{-1}, 1.98\times10^{-1}, 1.70\times10^{-1}$ for $\ell=10, 10^2, 10^3$. The relative weakness of the non-Gaussianities in the iSW-effect is related to fact that $\tau$ is dominated by contributions from fluctuations on comparatively large spatial scales, where structure formation is well described by linear theory.

In eqns.~(\ref{eqn_alpha_tau}) and~(\ref{eqn_alpha_gamma}), $C_{\tau\tau}(\ell)$ denotes the angular iSW power spectrum, $C_{\gamma\gamma}(\ell)$ the angular galaxy spectrum and $C_{\tau\gamma}(\ell)$ the cross-spectrum, respectively:
\begin{equation}
C_{\tau\tau}(\ell) = 
\int_0^{\chi_H}\dd\chi\:\frac{1}{\chi^2}
W_{\tau}^2(\chi)Q_{\tau}^2(a)\:P_{\varphi\varphi}(k=\ell/\chi),
\label{eqn_tautau_spectrum}
\end{equation}
\begin{equation}
C_{\gamma\tau}(\ell) = 
\int_0^{\chi_H}\dd\chi\:\frac{1}{\chi^2}
W_{\gamma}(\chi)W_\tau(\chi)Q_\gamma(a)Q_\tau(a)\:
P_{\delta\varphi}(k=\ell/\chi),
\end{equation}
\begin{equation}
C_{\gamma\gamma}(\ell) = 
\int_0^{\chi_H}\dd\chi\:\frac{1}{\chi^2}
W_{\gamma}^2(\chi)Q_{\gamma}^2(a)\:P_{\delta\delta}(k=\ell/\chi),
\label{eqn_gamgam_spectrum}
\end{equation}
with first-order time evolution $Q_\gamma(a)$ and $Q_\tau(a)$ of the respective fields (compare eqns.~\ref{eqn_time_evolution_delta} and~\ref{eqn_time_evolution_phi}) and the weighting functions $W_\tau(\chi)$ and $W_\gamma(\chi)$ (eqns.~\ref{eqn_w_tau} and~\ref{eqn_w_gamma}). $P_{\varphi\varphi}(k)=P_{\delta\delta}(k)/(d_Hk)^4$ and $P_{\delta\varphi}(k)=P_{\delta\delta}(k)/(d_H^2)$ are the potential and the density-potential cross spectrum, where the Hubble distance $d_H=c/H_0$ makes the wave vector $k$ dimensionless such that all three spectra have units of $(\mathrm{Mpc}/h)^{3}$.

\section{Detectability}\label{sect_fisher}

\subsection{Covariances}
In contrast to power spectra, the observed bispectra $\tilde{B}_{(abc)}(\ell_1,\ell_2,\ell_3)$ are unbiased estimates of the true bispectra $B_{(abc)}(\ell_1,\ell_2,\ell_3)$ in the case of Gaussian noise components,
\begin{equation}
\tilde{B}_{(abc)}(\ell_1,\ell_2,\ell_3)\simeq B_{(abc)}(\ell_1,\ell_2,\ell_3).
\end{equation}
Using a Gaussian approximation, the bispectrum covariance can be expanded into a product of three power spectra, and in the case of mixed bispectra, the covariance is a sum over all possible permutations of the observed cross-correlation $\tilde{C}_{\tau\gamma}(\ell)$ and the two auto-correlations $\tilde{C}_{\tau\tau}(\ell)$ and $\tilde{C}_{\gamma\gamma}(\ell)$. Specifically, the Gaussian approximation to the covariance $\mathrm{Cov}[B_{(abc)}(\ell_1,\ell_2,\ell_3),B_{(ijk)}(\ell_1^\prime,\ell_2^\prime,\ell_3^\prime)]$, which arises in the Wick-decomposition of the corresponding 6-point correlation function, assumes the shape \citep{1999ApJ...522L..21H, 2002PhRvD..66h3515H, 2003MNRAS.344..857T, 2004MNRAS.348..897T}
\begin{displaymath}
\mathrm{Cov}[B_{(abc)}(\ell_1,\ell_2,\ell_3),B_{(ijk)}(\ell_1^\prime,\ell_2^\prime,\ell_3^\prime)] =
\end{displaymath}
\begin{displaymath}
\qquad
\tilde{C}_{(ai)}(\ell_1)\tilde{C}_{(bj)}(\ell_2)\tilde{C}_{(ck)}(\ell_3)\:
\delta_{\ell_1\ell_1^\prime}\delta_{\ell_2\ell_2^\prime}\delta_{\ell_3\ell_3^\prime}\: +
\end{displaymath}
\begin{displaymath}
\qquad
\tilde{C}_{(ai)}(\ell_1)\tilde{C}_{(bk)}(\ell_2)\tilde{C}_{(cj)}(\ell_3)\:
\delta_{\ell_1\ell_1^\prime}\delta_{\ell_2\ell_3^\prime}\delta_{\ell_3\ell_2^\prime}\: +
\end{displaymath}
\begin{displaymath}
\qquad
\tilde{C}_{(aj)}(\ell_1)\tilde{C}_{(bi)}(\ell_2)\tilde{C}_{(ck)}(\ell_3)\:
\delta_{\ell_1\ell_2^\prime}\delta_{\ell_2\ell_1^\prime}\delta_{\ell_3\ell_3^\prime}\: +
\end{displaymath}
\begin{displaymath}
\qquad
\tilde{C}_{(aj)}(\ell_1)\tilde{C}_{(bi)}(\ell_2)\tilde{C}_{(cj)}(\ell_3)\:
\delta_{\ell_1\ell_2^\prime}\delta_{\ell_2\ell_3^\prime}\delta_{\ell_3\ell_1^\prime}\: +
\end{displaymath}
\begin{displaymath}
\qquad
\tilde{C}_{(ak)}(\ell_1)\tilde{C}_{(bj)}(\ell_2)\tilde{C}_{(ck)}(\ell_3)\:
\delta_{\ell_1\ell_3^\prime}\delta_{\ell_2\ell_1^\prime}\delta_{\ell_3\ell_2^\prime}\: +
\end{displaymath}
\begin{equation}
\qquad
\tilde{C}_{(ak)}(\ell_1)\tilde{C}_{(bj)}(\ell_2)\tilde{C}_{(ci)}(\ell_3)\:
\delta_{\ell_1\ell_3^\prime}\delta_{\ell_2\ell_2^\prime}\delta_{\ell_3\ell_1^\prime},
\label{eqn_covariance}
\end{equation}
and scales like $f_\mathrm{sky}^{-1}$ for an observation which covers a fraction of $f_\mathrm{sky}$ of the sky. $\tilde{C}_{(ai)}(\ell)$ corresponds to the observed CMB-spectrum $\tilde{C}_{\tau\tau}(\ell)$, if both indices are equal to one, the galaxy spectrum $\tilde{C}_{\gamma\gamma}(\ell)$ for both indices being equal to zero and the cross spectrum $\tilde{C}_{\tau\gamma}(\ell)$ for unequal indices:
\begin{equation}
\tilde{C}_{(ai)}(\ell) = 
\left\{
\begin{array}{ll}
\tilde{C}_{\tau\tau}(\ell),			& a+i=2,\\
\tilde{C}_{\tau\gamma}(\ell),			& a+i=1,\\
\tilde{C}_{\gamma\gamma}(\ell),		& a+i=0.
\end{array}
\right.
\end{equation}
I drop the negative sign of the cross spectrum $\tilde{C}_{\tau\gamma}(\ell)$ because there will be always an even number of cross spectra in the terms of the expression for the covariance~(\ref{eqn_covariance}).

\subsection{Noise sources}
\begin{figure}
\resizebox{\hsize}{!}{\includegraphics{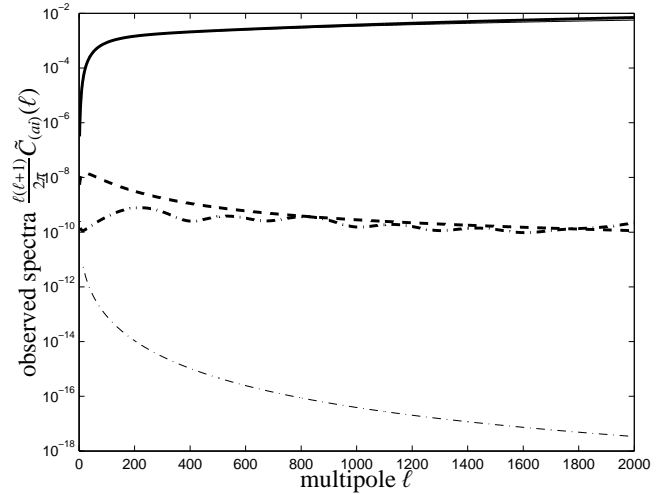}}
\caption{The observed spectra $\tilde{C}_{\gamma\gamma}(\ell)$ (thick solid line), $\tilde{C}_{\tau\gamma}(\ell)$ (thick dashed line) and $\tilde{C}_{\tau\tau}(\ell)$ (thick dash-dotted line), for PLANCK in conjunction with DUNE. In comparison, the corresponding noiseless spectra $C_{\gamma\gamma}(\ell)$ (thin solid line) and $C_{\tau\tau}(\ell)$ (thin dash-dotted line) are also plotted.}
\label{fig_noise_spectrum}
\end{figure}

The observed spectra $\tilde{C}_{\tau\tau}(\ell)$ and $\tilde{C}_{\gamma\gamma}(\ell)$ differ from the theoretical expectations by the primary CMB fluctuations, a Gaussian instrumental noise source $\sigma_\tau^2$ and a Gaussian beam $\beta(\ell)$ in case of the CMB observation, and by a Poissonian noise term $n^{-1}$ in case of the galaxy survey, assuming that the noise sources are mutually uncorrelated,
\begin{eqnarray}
\tilde{C}_{\tau\tau}(\ell)			& = & 
C_{\tau\tau}(\ell) + C_\mathrm{CMB}(\ell) + \sigma_\tau^2\:\beta^{-2}(\ell),\\
\tilde{C}_{\tau\gamma}(\ell)		& = & 
C_{\tau\gamma}(\ell), \\
\tilde{C}_{\gamma\gamma}(\ell)		& = & 
C_{\gamma\gamma}(\ell) + n^{-1}.
\end{eqnarray}
More specifically, the observational noise consists of these contributions:
\begin{enumerate}
\item{The Fourier-transform of a Gaussian beam is given by $\beta^{-2}(\ell)=\exp(\Delta\theta^2\ell(\ell+1))$. For the beam width I use the value $\Delta\theta=7\farcm 1$ corresponding to the $\nu=143~\mathrm{GHz}$ channels closest to the CMB emission maximum. The value $T_\mathrm{CMB}=2.725~\mathrm{K}$ for the CMB temperature is used when converting $w_T^{-1}=T^2_\mathrm{CMB}\sigma_\tau^2$ to the noise amplitude in the dimensionless temperature perturbation $\tau$, with $w_T^{-1} = (0.01~\umu\mathrm{K})^2$ \citep{1997ApJ...488....1Z}.}
\item{In addition, I generated a CMB temperature power spectrum $C_\mathrm{CMB}(\ell)$, equally scaled with the CMB temperature $T_\mathrm{CMB}=2.725$~K, with the Code for Anisotropies in the Microwave Background\footnote{\tt http://camb.info/} \citep[CAMB, ][]{2000ApJ...538..473L} for the fiducial $\Lambda$CDM cosmology. The noise contribution from the CMB-spectrum $C_\mathrm{CMB}(\ell)$ is the main difficulty in observing the iSW-bispectra, because it provides high values for the covariance at low multipoles $\ell$, and it by far dominates $\tilde{C}_{\tau\tau}(\ell)$, $C_\mathrm{CMB}(\ell)\gg C_{\tau\tau}(\ell)$ on the angular scales considered.}
\item{The Poissonian noise term in the galaxy counts is the inverse of the number density $n$ of objects per unit steradian. Table~\ref{table_survey} summarises properties of the DUNE main galaxy sample. The main advantage of DUNE is the large sky coverage and the high number of objects. For simplicity, I have assumed a constant (i.e. non-evolving) unit bias for the DUNE galaxy sample.}
\end{enumerate}
An overview over the observed spectra is given by Fig.~\ref{fig_noise_spectrum}, in comparison to the noiseless spectra: For the angular scales considered here, the Poisson noise term in the galaxy number counts is not yet an issue, the cross-spectrum and the intrinsic CMB spectrum are of similar magnitude, and at $\ell=10^3$ about a factor of $10^6$ larger than the noiseless iSW-spectrum.

\begin{table}\vspace{-0.1cm}
\begin{center}
\begin{tabular}{cccccc}
\hline\hline
$N$ 				& $\Delta\Omega$	& $f_\mathrm{sky}$	& $z_0$	& $b$	& $n$\\
\hline

$3.0\times10^9$		& $2\pi$			& 0.5				& 0.64 	& 1.0 	& $4.7\times10^8$  \\
\hline
\end{tabular}
\end{center}
\caption{Properties of the DUNE survey: total number $N$ of objects, solid angle $\Delta\Omega$ covered (in radians), sky fraction $f_\mathrm{sky}$, redshift parameter $z_0$, bias $b$ and density per unit steradian $n$.}
\label{table_survey}
\end{table}

\subsection{Signal to noise ratios}
\begin{figure}
\resizebox{\hsize}{!}{\includegraphics{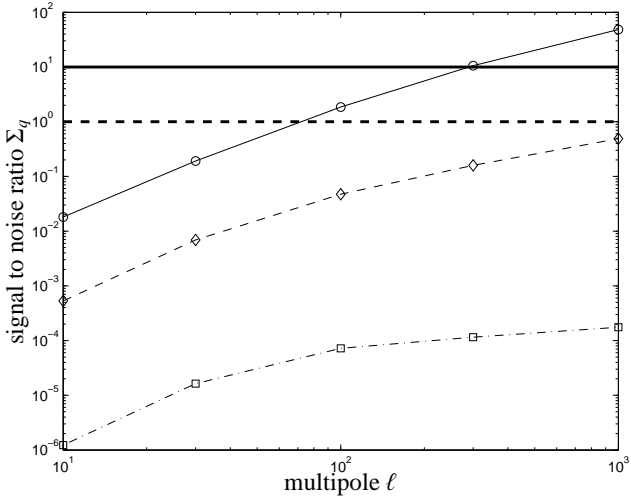}}
\caption{Signal to noise ratio $\Sigma_q$ of the bispectrum $\bra\tau^q\gamma^{3-q}\ket$ as a function of maximum multipole moment $\ell_\mathrm{max}$: $\Sigma_0$ (solid line), $\Sigma_1$ (dashed line) $\Sigma_2$ (dash-dotted line), for cross-correlating the DUNE galaxy sample with PLANCK CMB data. The horizontal lines mark the confidence levels $1\sigma$ (thick dashed line) and $10\sigma$ (thick solid line).}
\label{fig_signal2noise}
\end{figure}

The signal to noise ratio $\Sigma_q$ (assuming a Gaussian likelihood) for a measurement of the bispectrum $\bra\tau^q\gamma^{3-q}\ket$ is given by \citep{2000PhRvD..62d3007H, 2003MNRAS.344..857T, 2004MNRAS.348..897T, 2006MNRAS.366..884D}:
\begin{equation}
\Sigma_q^2 = 
\sum_{\ell_p=\ell_\mathrm{min}}^{\ell_\mathrm{max}}
\sum_{\ell_p^\prime=\ell_\mathrm{min}}^{\ell_\mathrm{max}}
\sum_{(abc) \atop a+b+c=q}\sum_{(ijk) \atop i+j+k=q}
B_{(abc)}(\ell_p)\: Q_{abc}^{ijk}(\ell_p,\ell_p^\prime)\: B_{(ijk)}(\ell_p^\prime),
\label{eqn_s2n}
\end{equation}
with $\ell_p\in\left\{\ell_1,\ell_2,\ell_3\right\}$ and $\ell_p^\prime\in\left\{\ell_1^\prime,\ell_2^\prime,\ell_3^\prime\right\}$. For convenience, I abbreviated the inverse covariance of the bispectrum $B_{(abc)}(\ell_1,\ell_2,\ell_3)$,
\begin{equation}
Q_{abc}^{ijk}(\ell_p,\ell_p^\prime) =
\left(
\mathrm{Cov}[B_{(abc)}(\ell_1,\ell_2,\ell_3), B_{(ijk)}(\ell_1^\prime,\ell_2^\prime,\ell_3^\prime)]
\right)^{-1}.
\end{equation}
Like \citet{2003MNRAS.344..857T, 2004MNRAS.348..897T}, I use binned summations in the multipoles $\ell_1$ and $\ell_2$, but carry out an unbinned summation in $\ell_3$ in order to account for the vanishing Wigner-$3j$ symbol if $\sum_p\ell_p$ is an odd number and for the sign change of the Wigner-$3j$ symbol depending on whether $\sum_p\ell_p~\mathrm{modulo}~4$ vanishes or not. Fig.~\ref{fig_signal2noise} shows the cumulative signal to noise ratio of the measurement of $B_q$ as a function of maximum multipole order $\ell_\mathrm{max}$. In the following I will use $\Delta\ell_1=\Delta\ell_2=30$ for the largest $\ell_\mathrm{max}$-values considered, and the summation is carried out starting from $\ell_\mathrm{min}=10$. When computing the covariance of the bispectrum $\bra\tau^q\gamma^{3-q}\ket$, one needs to make sure that each term in the summation contains exactly $2q$ iSW-fields $\tau$, which is achieved by restricting the summation over all possible permutations to those terms which satisfy the conditions $a+b+c=q$ and $i+j+k=q$. 

As shown by Fig.~\ref{fig_signal2noise}, the cumulative signal to noise ratio $\Sigma_q$ increases slowly with increasing multipole order $\ell$. The galaxy bispectrum $\bra\gamma^3\ket$ should be measurable with $3\sigma$ significance on degree angular scales. The bispectrum $\bra\tau\gamma^2\ket$, sadly, only reaches a significance level of $\simeq0.4\sigma$ up to the scale $\ell=10^3$, and one would need to carry on the measurement to angular scales of $\simeq10^4$ for a detection significance of $3\sigma$. These angular scales are beyond PLANCK's resolution limit and other secondary anisotropies such as the Sunyaev-Zel'dovich effect or the contribution to point sources becomes important. Furthermore, the second order perturbation used in this work would be no longer applicable. At $\ell=10^3$, the $\Sigma_0$ is roughly two orders of magnitude larger than $\Sigma_1$, which in turn is slightly more than three orders of magnitude larger than $\Sigma_2$. $\Sigma_2$ is separated from $\Sigma_3$ by four orders of magnitude. This behaviour reflects the increasingly higher powers of the primary CMB power spectrum $C_\mathrm{CMB}(\ell)$ which enters the covariance as a noise source.

Signal to noise ratios $\Sigma_q$ for measurements of all three-point functions by correlation with the DUNE main galaxy sample are compiled in Table~\ref{table_signal2noise}, where the summation was extended to the maximum multipole moment $\ell=2\times10^3$. The value of the signal to noise ratio $\Sigma_1$ for a measurement of the bispectrum $\bra\tau\gamma^2\ket$ amounts to $0.6\sigma$ considering triangle configurations up to PLANCK's resolution limit, which is not large enough to be realistically detected, because the confidence level would only be $\mathrm{erf}(0.6/\sqrt{2})\simeq0.45$. As a consistency check, I also derived the signal to noise ratio $\Sigma_3$ for measuring the bispectrum $\bra\tau^3\ket$, and obtain a value of $\Sigma_3\simeq10^{-8}\sigma$, which is quite comparable to the estimate of \citet{2002PhRvD..65h3518C}, keeping in mind that we use different cosmological models and a different perturbative approach.

\begin{table}\vspace{-0.1cm}
\begin{center}
\begin{tabular}{lll}
\hline\hline
$\bra\gamma^3\ket$			& $\bra\tau\gamma^2\ket$	& $\bra\tau^2\gamma\ket$ \\
\hline
$\Sigma_0 = 56.29$			& $\Sigma_1 = 0.593$ 		& $\Sigma_2=1.667\times10^{-4}$ 	\\
\hline
\end{tabular}
\end{center}
\caption{Cumulative signal to noise ratios $\Sigma_q$ for measurements of the three-point correlation functions $\bra\tau^q\gamma^{3-q}\ket$, $q=0,1,2$, for PLANCK CMB data in cross-correlation with the DUNE survey, up to PLANCK's resolution limit $\ell_\mathrm{max}=2\times10^3$.}
\label{table_signal2noise}
\end{table}

\subsection{Degeneracies}\label{sect_degeneracy}
The Fisher matrix \citep{1997ApJ...480...22T}, which describes the logarithmic decrease in likelihood if the cosmological parameters $x_\mu$ move away from their fiducial values, is defined in analogy to eqn.~(\ref{eqn_s2n}):
\begin{equation}
F^{(q)}_{\mu\nu} = 
\sum_{\ell_p=\ell_\mathrm{min}}^{\ell_\mathrm{max}}
\sum_{\ell_p^\prime=\ell_\mathrm{min}}^{\ell_\mathrm{max}}
\sum_{(abc) \atop a+b+c=q}\sum_{(ijk) \atop i+j+k=q}
\frac{\partial B_{(abc)}(\ell_p)}{\partial x_\mu} 
Q_{abc}^{ijk}(\ell_p,\ell_p^\prime)
\frac{\partial B_{(ijk)}(\ell_p^\prime)}{\partial x_\nu},
\label{eqn_fisher}
\end{equation}
but the derivation of parameter degeneracies would not yield competitive parameter bounds, given the low values for the signal to noise ratio $\Sigma_q$. Quite generally, all bispectra are proportional to $\sigma_8^4$ because of the perturbative evaluation to second order, and $\propto\Omega_m^q$ because the matter density plays the role of a coupling strength in the line of sight expression for the iSW perturbation $\tau$. The bispectra with small $q$ are more strongly influenced by the Hubble parameter $h$ via the shape parameter $\Gamma=\Omega_m h$ because the strong weighting $\propto 1/k^4$ of the potential power spectrum dampens the sensitivity. Concerning the line of sight integrated quantities the scale factor $a$ at which equality $\Omega_m(a)=\Omega_\Lambda(a)$ is reached, plays an important role, because at the corresponding redshift most of the iSW-signal is created. Furthermore, models with small average dark energy equation of state parameter $w$ give rise to an iSW-effect at higher redshifts compared to $\Lambda$CDM.

\section{Summary}\label{sect_summary}
The aim of this study is an investigation of the detectability of the nonlinear iSW-effect using mixed bispectra of the form $\bra\tau^q\gamma^{3-q}\ket$, $q=0,1,2$, between the galaxy density $\gamma$ and the iSW-temperature perturbation $\tau$. The bispectra were consistently derived in second order perturbation theory. I investigated the time evolution of the souce terms, the configuration dependence of the bispectra, the achievable signal to noise ratio in a measurement cross-correlating PLANCK and DUNE data, and their parameter sensitivity.
\begin{enumerate}
\item{The nonlinear iSW-effect has a different parameter dependence compared to the linear iSW-effect, as it is sensitive to the derivative $\dd(D_+^2/a)/\dd a$ instead of $\dd(D_+/a)/\dd a$. Particularly, it does not vanish in SCDM-models, where $\Omega_m\equiv 1$ and consequently $D_+(a)=a$.}
\item{I employed second order hyper-extended perturbation theory for deriving the the bispectra, and used a Gaussian approximation for describing the covariance of the measurements. As noise sources, I considered the intrinsic CMB fluctuations, pixel noise and a Gaussian beam for the CMB observation, and a Poissonian noise term in the galaxy density. To this point, I worked with a constant linear biasing model for relating the fluctuations in galaxy number density to those of the dark matter density.}
\item{The configuration and scale dependence of the mixed bispectra reflects the interplay between the correlation length of the density field and the much larger correlation length of the gravitational potential.}
\item{I computed the signal to noise ratio for the measurements of bispectra $\bra\tau^q\gamma^{3-q}\ket$, $q=0,1,2$, with a Gaussian approximation to the covariance, in which the intrinsic CMB fluctuatios are the most important noise source which make the bispectra difficult to observe. Values for the cumulative the signal to noise ratio $\Sigma_1\simeq0.6\sigma$ for the measurement of the bispectrum $\bra\tau\gamma^2\ket$ are obtained for cross correlating PLANCK data with the DUNE main galaxy sample, up to PLANCK's resolution limit at $\ell=2\times10^3$, where the dominating noise source are the primordial CMB fluctuations.}
\item{An algorithm for evaluating 3-point correlation function would need to be implemented for PLANCK data processing. This is most likely an algorithm operating in harmonic space instead of real space, because the evaluation of the 3-point correlation functions scales as $N_\mathrm{pix}^3$ where $N_\mathrm{pix}$ is the number of pixels, which becomes prohibitive at $N_\mathrm{pix}\simeq 10^7$ , corresponding to multipole orders of $\ell=1000$. From the observational point of view, the influence of unresolved microwave point sources or the kinetic Sunyaev-Zel'dovich effect, which are equally associated with overdensities in the LSS, on a measurement of the iSW-temperature fluctuation is yet unquantified.}
\end{enumerate}

Future studies will treat three-point correlation functions of the type $\bra\tau^q\kappa^{3-q}\ket$ between the iSW temperature perturbation $\tau$ and weak gravitational lensing convergence $\kappa$, which would be attractive because lensing measures directly the fluctuations in the dark matter density, without uncertainties related to bias and bias evolution. In comparison to $\bra\tau^q\gamma^{3-q}\ket$, the bispectrum $\bra\tau^q\kappa^{3-q}\ket$ would measure the dark energy properties at lower redshifts. For the DUNE sample, the constraints would come from redshifts of $z\simeq0.4$, compared to $z\simeq0.9$, with more strongly evolved nonlinear structures. A third point in favour of gravitational lensing is the smaller sampling noise in the galaxy ellipticity $\sigma_\epsilon^2/n\simeq0.1/n$ compared to $1/n$ in the galaxy density.

\section*{Acknowledgements}
My work is supported by an STFC postdoctoral fellowship. I would like to thank Alexandre Refregier, Nabila Aghanim and Marian Douspis for providing the redshift distribution of the DUNE galaxy sample. I very much appreciate comments on the nonlinear iSW-effect from Chema Diego, Patricio Vielva-Mart{\'i}nez, Carlos Hern{\'a}ndez-Monteagudo and Rob Crittenden.

\bibliography{bibtex/aamnem,bibtex/references}
\bibliographystyle{mn2e}

\appendix

\bsp

\label{lastpage}

\end{document}